\documentstyle[12pt,psfig]{article}
\begin{document}
\sloppy

\begin{center}

{\Large{\bf
EXPERIMENTAL INVESTIGATION OF CHANGES IN $\beta$-DECAY
COUNT RATE OF RADIOACTIVE ELEMENTS
}}
\vskip20pt

                          Yu.A. BAUROV\footnote{alex@theor.phys.msu.su}

{\it     Central Research Institute of Machine Building,

141070, Korolyov,  Moscow Region, Russia}
\vskip10pt

                Yu.G. SOBOLEV\footnote{ak@sky.chph.ras.ru}, V.F. KUSHNIRUK and E.A. KUZNETSOV

{\it Flerov Laboratory of Nuclear Reactions (FLNR),

Joint Institute for Nuclear Research,

141980, Dubna, Moscow Region, Russia}
\vskip10pt

                          A.A. KONRADOV\footnote{sobolev@main1.jinr.dubna.su}

{\it Russian Academy of Sciences, Institute of Biochemical Physics,

 117977, Moscow, Russia}
\vskip10pt

\vskip25pt
 ABSTRACT
\end{center}
{\footnotesize
The experimental data on continuous investigation of changes in
$\beta$-decay count rate of ${}^{137}Cs$ and ${}^{60}Co$ from 9.12.98 till
30.04.99, are presented. The 27-day and 24-hour periods in these
changes, inexplicable by traditional physics, have been found.}
\vskip25pt
{\bf 1. Introduction }
\vskip10pt
In Refs. [1-3], periodic variations in $\beta$-decay rate of ${}^{60}Co$,
${}^{137}Cs$, and ${}^{90}Sr$, have been first discovered. An analysis of the
24-day period in $\beta$-decay of radioactive elements as well as of
the daily rotation of the Earth in various seasons of the year
has led to selection of some spatial direction characterized by
the fact that near the points of the Earth's surface where the
latitude tangent line to a parallel passes through this
direction, the decay count rate of radioactive elements changes.
The main drawback of the experiments [1-3] was that their final
results gave no possibility to clearly understand what was an
effect of the "internal life" of the setup itself and what was
due to the phenomenon of interest. In addition, the duration of
these experiments was no more than three weeks, which did not
allow to analyze long-period harmonics.

The aim of this paper is to find an answer to the above questions,
using measurements of flux of $\gamma$-quanta in the process of
$\beta$-decay of radioactive elements as in ref. \cite{1}.
\vskip10pt
{\bf 2. The diagram of the setup.}
\vskip10pt
The experimental setup (Fig.1) consisted of three scintillation
detectors, two of them being standard spectrometric
scintillation detecting units {\bf BDEG2-23} on the basis of {\bf NaI}($Tl$)-scintillator
(63mm in diameter, 63mm in height) and {\bf FEU-82}
photomultiplier ({\bf PM}) with standard divider. One of these units
was used to indicate the background radiation, and the second
one was to present $\gamma$-radiation of ${}^{137}Cs$. The third detector was
a {\bf BGO}-scintillator (46mm in diameter, 60mm in height) and a
{\bf FEU-143} photomultiplier with standard divider. This detector was
used to display the $\gamma$-radiation of a ${}^{60}Co$-source.

To diminish the influence of magnetic fields on the {\bf PM}s, 
the detectors were placed into protecting screens made as cilynders
from ten sheets of annealed permalloy 0.5 mm in thickness. The
internal diameter of the cylinders was equal to 10 cm, and the
height was 70 cm.

The detectors were placed in such a manner
that the photocathodes of {\bf PM}s were at a distance of one-half of
height of the cylinder. The $\gamma$-sources were placed just on the
end face surface of the scintillators through the center of the
input window.
All the detectors and the temperature-sensitive-element were positioned inside the
metallic cube ($40\times40\times50$ cm$^3$) used as an additional magnetic
shield. The thickness of the steel walls of the cube was
equal to 3 mm. The detectors with $\gamma$-sources were surrounded by
lead protection 5 cm in thickness.
\vskip10pt
{\bf 3. The system of registering experimental information.}
\vskip10pt
The system of registering information consisted of two subsystems.
The first one was designed for accumulating information on the
counting rate in ten-second intervals from 
scintillation detectors as well as on the temperature,
power-source voltages (high voltage of {\bf PM}s, voltage of {\bf CAMAC}
$= 6V, 24V$) and impulse noise of crate power supply. The second
subsystem of information storage were made to record "marked"
energetic distributions from scintillation detectors for the
purpose of checking the stability of their amplitude
distribution parameters (stability of discriminator thresholds,
shape of amplitude distributions etc.).
\vskip5pt
{\it 3.1. The spectrometric sections.}
\vskip5pt
The set included three identical spectrometric registering
sections (see Fig.1). Each section consisted of a preamplifier
({\bf PA})-emitter follower matching the impedances, spectrometric
amplifier ({\bf AFA}) with active filters having shaping time
constants $T_{int} = T_{dif} = 0.25 \mu s$ \cite{4}, and system of fast discriminators
({\bf FD}) of negative output signals of the amplifier and counters of
gated pulses ({\bf SC}). In addition, the positive output signal of
the spectrometric amplifier {\bf AFA} from each registering channel
was fed to analog-to-digital converters ({\bf ADC}) \cite{5} placed in a
separate crate. To increase reliability of spectrometric
section, all variable resistors in which, as almost 20-year
operating experience has shown, sometimes the contact arm faults
take place (when continuously adjust amplification in {\bf AFA} and
the threshold in {\bf FD}), were replaced by the fixed resistors.
The {\bf PM}s 
of all sections had the general high-voltage power supply.
\vskip5pt
{\it 3.2. The system of monitoring and recording parameters.}
\vskip5pt
In long-term experiments, the most important requirement upon
the measuring system is the possibility of continuous control
over its parameters as for detecting non-stable elements, units,
and connections, so to refine possible correlations of
measurable quantities with the environmental parameters.  The
experimental setup was powered from separate terminals of
distributing board for diminishing the possible influence of
additional parallel loads in the power network.

To monitor the temperature of the environment, a thermometric channel with
high-sensitive temperature element and amplifier module was
used. This element was made on the basis of assembly of
semiconductor diodes with the summary thermoelectric coefficient
about 10 mV/degree. The amplifier module gave stable bias
current for the temperature-sensitive element and additionally
amplified the signal up to the summary termoelectric coefficient
of the measuring channel of 100 mV/degree. In the same
module a transformer of voltage from high-voltage power supply of the
scintillation units into low voltage for {\bf 8ADC} (see below) was
arranged. The transformation ratio was about 3.3 V/kV.

In the measuring crate with the counters, amplifiers {\bf AFA}, and the
amplifier module of the thermometric channel, we have placed
also a multichannel amplitude-to-digital converter {\bf 8ADC} for
measuring high voltage ({\bf HV}) of the scintillation detectors and
monitoring the secondary power voltages $\pm 24V, \pm 6V$ of the crate
{\bf CAMAC} itself, as well as a special module to register the
impulse noise of these secondary power sources. Any impulse
input in the crate power line with an amplitude more than 10mV
recorded "1" into the corresponding information bit of the word
register of module data. The frequency spectrum of recorded
impulse signals extended from tens Hz to several MHz. Thus we
recorded impulse noise of the crate along with monitoring levels
of constant high voltages of the power source of the
scintillation units as well as low voltages of the crate power
sources.

  The start of measuring cycle and quantization of
exposure time in the first recording subsystem were organized by
a "Master-Trigger" {\bf MT1}. It comprised a pulser 
with quartz stabilization of frequency of output pulses ({\bf QUARTZ}) and a
scaling circuit.  Each cycle of measurements in the experiment
started with generation of a ten-second exposure signal ({\bf GATE})
by the unit {\bf MT1}. This signal opened all counters of the setup
({\bf SC1-SC6}).  After the ten-second signal of exposure of the
counters, {\bf MT1} elaborated a signal "{\bf LAM}" for the controller {\bf CC1}
of the measuring crate to organize a cycle of interrogation of
the crate recorders and transmission of date to the storage {\bf PC1}.
The data file, transmitted to {\bf PC1} in each interrogation cycle,
included the following data words:
\vskip5pt
- \hbox to 12.5cm{the number of readings in the counters {\bf SC1-SC6} \hfil   $8\times16$ bits},

- \hbox to 12.5cm{the codes of voltages of {\bf CAMAC} sources $\pm 6V, \pm 24V$ \hfil $4\times15$ bits},

- the code of voltage of the high-voltage power source

\hbox to 13.5cm{\hspace{.9cm}of the scintillation  units \hfil 15 bits,}

- \hbox to 12.5cm{the code of the recorder of impulse noise \hfil 4 bits.}
\vskip5pt
\noindent
The 15-digit codes with {\bf 8ADC} contained 12 bits of the voltage code and 3 bits of
the channel number.

The characteristics of the sections:
\vskip5pt
a) \it Sensitivity \rm (the exposure time 10 s with an accuracy of $10^{-6}$ s):
\begin{center}
\begin{tabular}{lcl}
"$\pm 6V$" & = & 5mV per channel;\\
"High-voltage power"& = &750 mV per channel;\\
"$\pm 24V$" & = & 12.5mV per channel; \\
"Temperature" & = & $1^\circ$ for 40 channels.
\end{tabular}
\end{center}

b) \it Thresholds \rm of the section ${}^{137}Cs$ {\bf NaI}($Tl$) (calibration against $\gamma$-lines 662 keV, 1173 keV, 1332 keV):
\begin{center}
\begin{tabular}{lcr}
the "low" threshold & = & 7 keV;\\
the threshold "under the peak" & = & 425 keV;\\
the threshold "on the peak" & = & 657 keV.
\end{tabular}
\end{center}

c) \it Thresholds \rm of the background section {\bf NaI} ($Tl$)
(calibration against $\gamma$-lines 662 keV, 1173 keV, 1332 keV):

\begin{center}
\begin{tabular}{lcr}
the "low" threshold & = & 11 keV.
\end{tabular}
\end{center}

d) \it Thresholds \rm of the ${}^{60}Co$ {\bf BGO}-section
(calibration against $\gamma$-lines 662 keV, 1173 keV, 1332 keV):

\begin{center}
\begin{tabular}{lcr}
the "low" threshold & = & 35 keV;\\
the threshold "under the peak" & = & 745 keV.
\end{tabular}
\end{center}

The start of measurements in the second recording subsystem
was organized by the "Master-Trigger" {\bf MT2} from any signal of the
discriminators {\bf FD1-FD6} (chosen by the experimenter by way of
switching from one channel to another in the module {\bf M}). The unit
{\bf MT2} opened by its {\bf GATE}-pulse the amplitude-code converters 
{\bf ADC1-ADC3}, "spectrum mark" counter {\bf SC7}, and triggered the cycle
of recording information into the storage computer {\bf PC2} after
the time of amplitude-digital code transformation.  The
{\bf GATE}-pulses from the {\bf MT1} unit of the first recording subsystem
were fed to the counter {\bf SC7} input. Thus the counter gave
information on numbers of ten-second exposure intervals of the
first subsystem. This allowed to perform analysis (in "off-line"
mode) of the amplitude distribution parameters  of  the chosen
channel of recording in any combination of ten-second exposures.

\vskip10pt
{\bf 4. The basic results of the experiment. Brief discussion.}
\vskip10pt

The long-term dynamics of the radioactive decay of  ${}^{137}Cs$ and
${}^{60}Co$ over the period from 9 December 1998 till 30 April 1999,
was measured. The above described setup made it possible to
perform precision measurements with monitoring parameters of the
system at the different discrimination thresholds of decay
energy. The spectra in the channels for ${}^{137}Cs$ are presented in
Figs.2-4 with the corresponding thresholds. As an example, in
Fig.5 the results of measurements over two-week time interval at the
end of March, 1999, are shown, for 7 main variants of channels.

\begin{tabular}{cc}
Variants of channels& Measurements\\
1.& {\bf BGO}, the threshold of Fig.3-type;\\
2.& {\bf BGO}, the threshold of Fig.2-type;\\
3.& {\bf NaI}$^1$  with the threshold in Fig.3;\\
4.& {\bf NaI}$^1$  with the threshold in Fig.4;\\
6.& {\bf NaI}$^1$  with the threshold in Fig.2;\\
12.& Internal temperature of the setup;        \\
13.& High voltage ({\bf HV}) in channels {\bf NaI}$^{1,2}$  and {\bf BGO}.
\end{tabular}
\vskip5pt
In the present paper we shall analyze
only the channel 6 corresponding to the minimum threshold of
discrimination at which only low-energetic noise component was
cut off, and the channel 12. From Fig.5 one can conclude that
the channel with the low discrimination was the most stable
though with a remarkable local dispersion: the data densely fill
a relatively broad band.

The starting series have more than
$1.2\times 10^6$ points in summary length over the whole time interval of
observation.  Each point corresponds to a ten-second interval of
decay number accumulation. Hence, the summary duration of
continuous measurements was $\sim 3347$ hours, or $\sim 140$ days.

When analyzing the periodical structure of the series we were
interested in periods no shorter than several hours.
In Fig.6 the results of normalized the starting series
(i.e. reducing to the interval [0,1]) averaging over one-hour period,
are given. With such hourly averaging, the "fast" component of
dispersion disappeared, and the slow dynamic of the process was
clearly  seen. It is also clear from the Figure that the
temperature inside the setup varied in antiphase with the
count rate. 
This is well seen in the whole
long series and, partially, in Fig.5. The cross-correlation
function of these two series has a sharp minimum approaching -0.95
at the zero lag. This allowed us to take into account the
temperature dependence of count rate measurements by way of
simple addition of two normalized series.  The Fourier-analysis
(fast Fourier transformation - {\bf FFT}) of the final
temperature-compensated series have revealed two distinctly
distinguishable periods. In Fig.7 a pronounced 27-day period is
seen that may be caused, for example, by the influence of the
Sun's rotation around its axis (the synodic period of the Sun's
rotation relative to the Earth is equal to 27.28 days). In the
hour-scale of the periods in Fig.8, a 24-hours period is well
marked. It should be emphasized that this daily period is absent
in the spectrum of the dynamic of the temperature itself (see
Fig.9) and is found only in the dynamic of the radioactive
decay, so that it can have an external cosmic reason, too.

Now let us consider the statistics of extremum values of the series
of measurements. Evaluate more accurately the extent of
nonuniformity of distribution of extremum values for the
starting (10-second) series of observations in the low-threshold
channel over the time of astronomical day. This procedure was
described earlier \cite{3}. Here we give its brief presentation.

Under an extremum we mean here a value for which the modulus of
difference with the average for the whole series is no lesser
then two standard deviations. Ascribing to each extremum value
that instant of day time at which this extremum was observed we
shall have the resulting set of time instants in the interval
from 0 till 24 hours when "jumps beyond two sigmas" were
measured. The "null hypothesis" consists in that the extremum
events occur with equal frequency in any time of day, i.e. the
distribution of these instances is uniform over the day cycle.
The hypothesis of uniformity of distribution can now be
validated, for example, by the Kolmogoroff-Smirmoff's test. In
Figs.10 and 11 the results of computations are presented. The
time of day laid as abscissa is expressed in degrees ($0-360^\circ$).

As a reference point, the time from beginning of observations is
taken (the start on 9 Dec. 1998 at $23^h$ of astronomical time -
the local time. The whole time of experiment was divided for
this analysis into exact decades in days). The values of
difference between the sample and uniform distribution functions
for each moment of day time are plotted as ordinates (in
degrees). With the dashed line the confidence levels of
Kolmogoroff-Smirnoff's criterion ($P<0.05$) are shown. An exit
beyond these limits denotes a significant difference of the
distribution from the uniform one, and the maximum point
indicates the time of day (phase) when this nonuniformity  was
maximum. In Fig.10 the results for the maximum values, and in
Fig.11 for minimum values are given.

The existence of reliable nonuniformity denotes presence of a daily period in the
statistics of extremum values of the radioactive decay. A
knowledge of phase (moment of maximum nonuniformity) as well as
relation to the absolute time (from the beginning of the series)
allows us to determine possible cosmic references connected with
such a nonuniformity.

The analysis of the extremum jumps have
shown that they overlie tangent lines to the Earth's parallels
making an angle of $\pm(35^\circ-45^\circ)$ with the direction having the
right ascension coordinate $\alpha \approx 275^\circ$ that insignificantly ($\sim 5^\circ$)
differs from the direction fixed in Refs.[1-3,6]. It should be
noted also that, as background measurements in the channel 5
have shown (the flux of particles in this channel was no more
then 50 particles per ten second), the oscillations of the
background (due to the smallness of its flux as compared with
that ($\sim 300$ per second) going beyond the scope of $2\sigma$) by no
means could influence on the distribution of  temporal
coordinates of the extremum points.

\begin{figure}[hb]  
\centerline{\psfig{figure=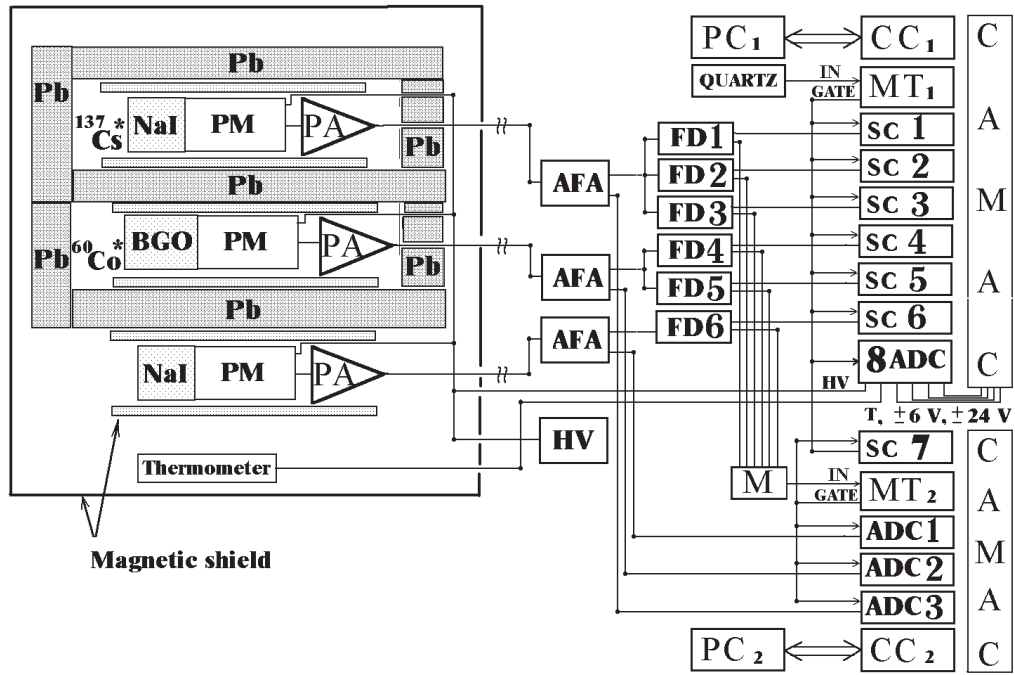,angle=0,height=90mm,width=137mm}}
\vspace{5mm}
\caption{
{\bf NaI} - scintillation detector {\bf NaI}(Tl); 
{\bf BGO} - scintillation detector (bismuth germanat); 
{\bf Pb} - lead shield 5cm in thickness;
{\bf PM} - photomultiplier;
{\bf PA} - preamplifier;
{\bf AFA} - active filter amplifier;
{\bf FD}(1-6) - fast discriminators;
{\bf SC} - scaler;
{\bf CC} - crate controller;
{\bf PC} - storage computer;
{\bf 8ADC} - analog to digital converter;
{\bf MT} - master yrigger;
{\bf QUARTZ} - guartz generator;
{\bf M} - multiplexer;
{\bf HV} - high voltage power supply.
}
\end{figure}

\begin{figure}  
\centerline{\psfig{figure=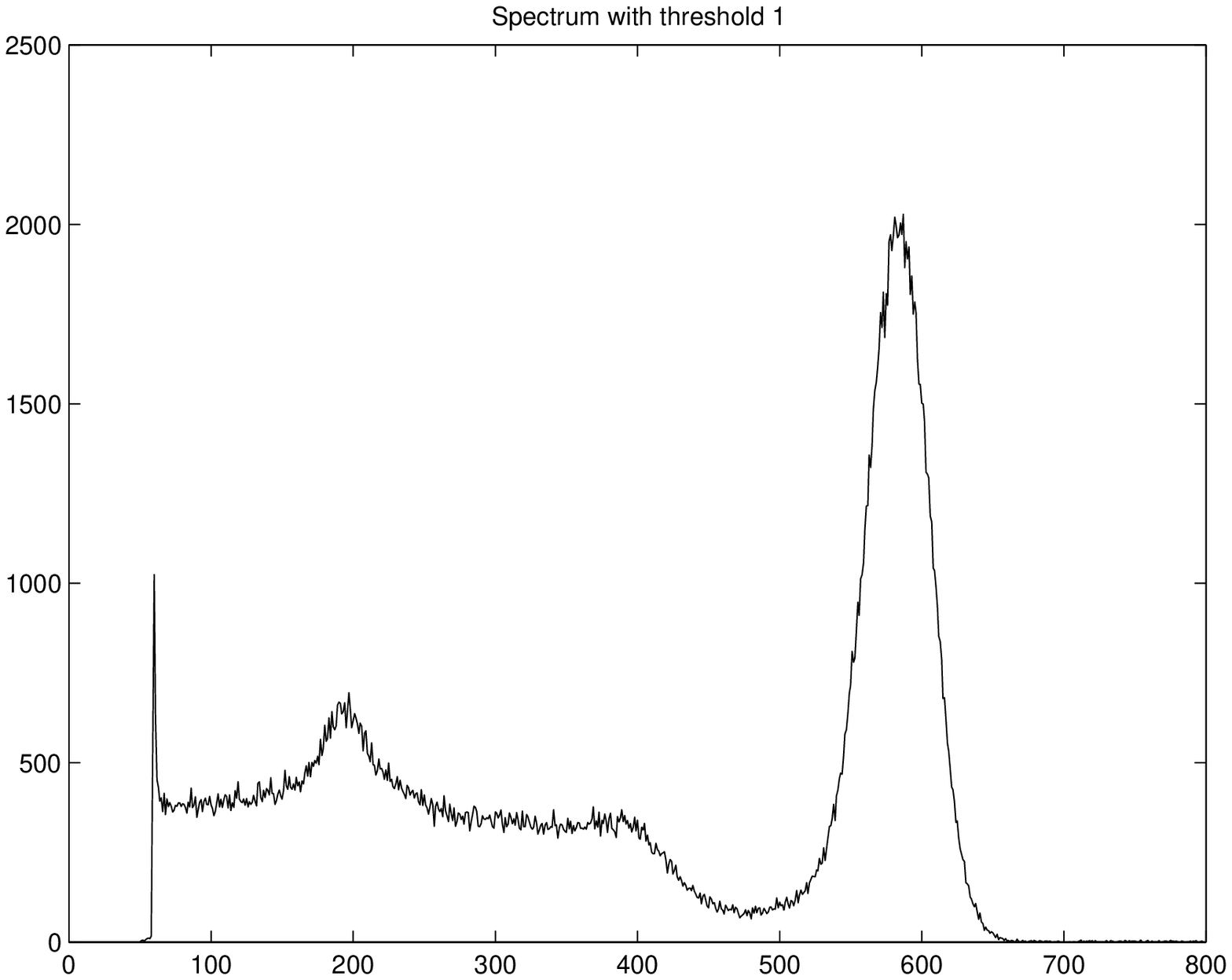,angle=0,height=60mm,width=60mm}}
\vspace{5mm}
\centerline{\hbox{Fig.2.}}
\end{figure}

\begin{figure}[hb]  
\centerline{\psfig{figure=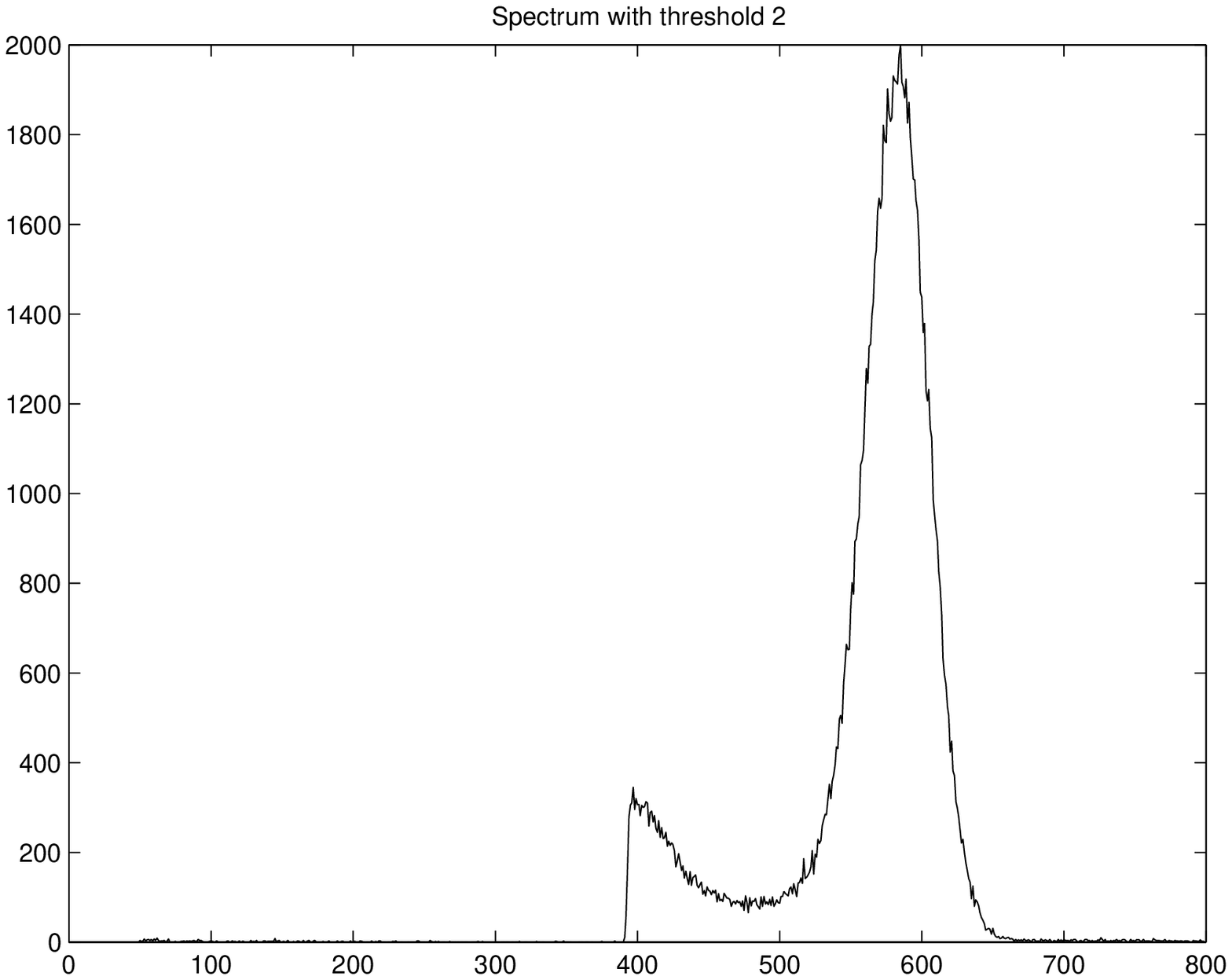,angle=0,height=60mm,width=60mm}}
\vspace{5mm}
\centerline{Fig.3.}
\end{figure}

\begin{figure}[hb]  
\centerline{\psfig{figure=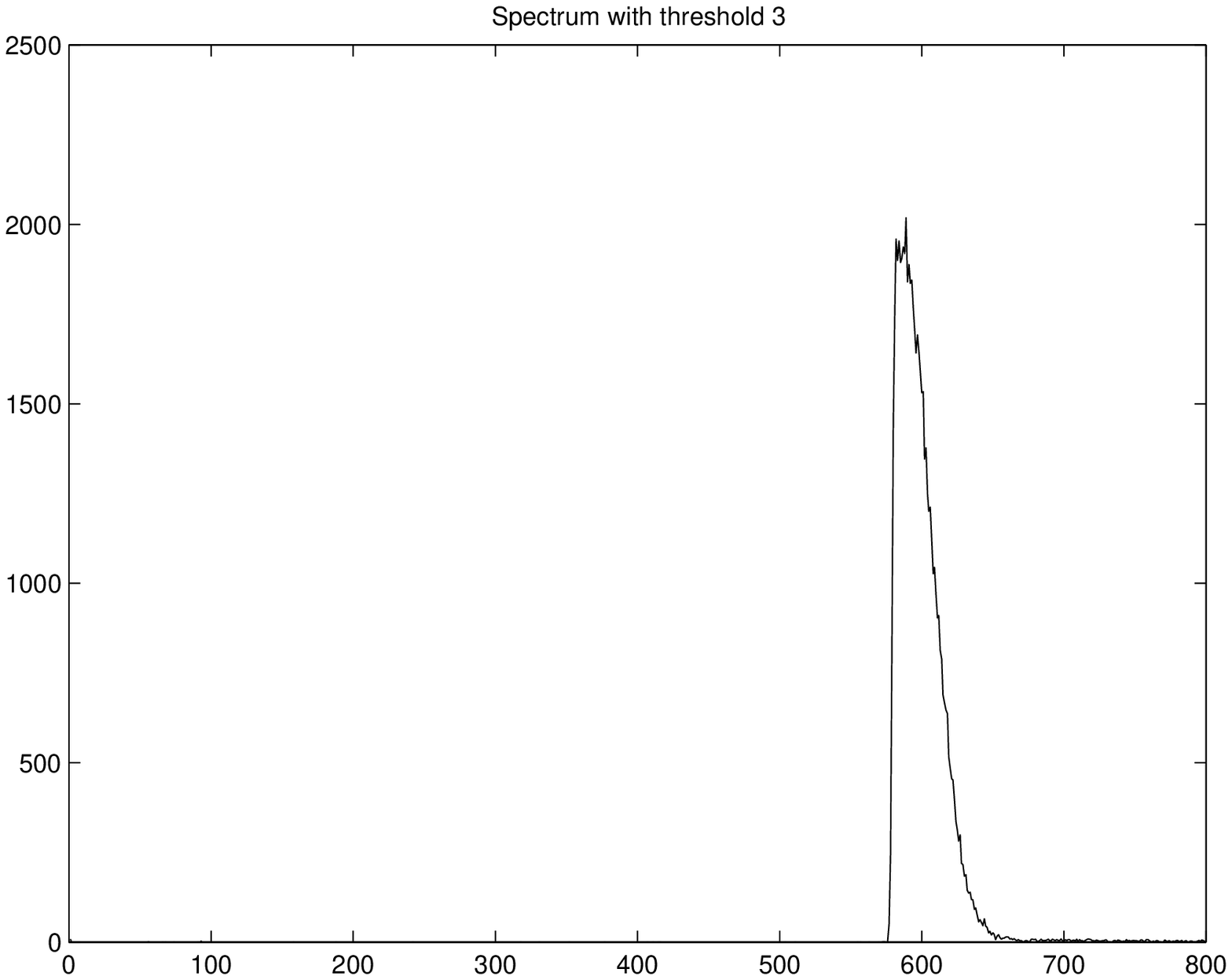,angle=0,height=60mm,width=60mm}}
\vspace{5mm}
\centerline{Fig.4.}
\end{figure}

\begin{figure}[hb]  
\centerline{\psfig{figure=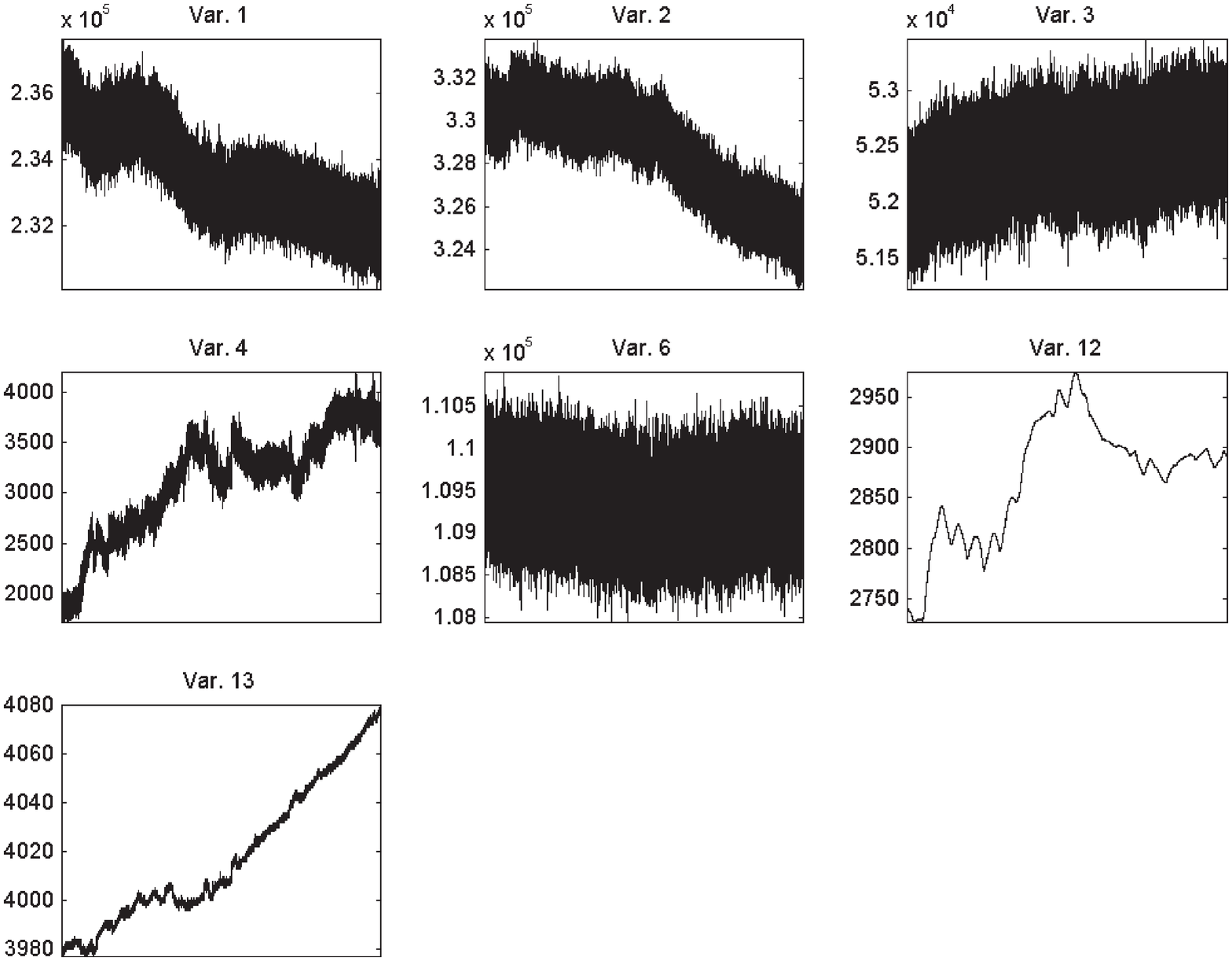,angle=0,height=80mm,width=100mm}}
\vspace{5mm}
\centerline{Fig.5.}
\end{figure}

\begin{figure}[hb]  
\centerline{\psfig{figure=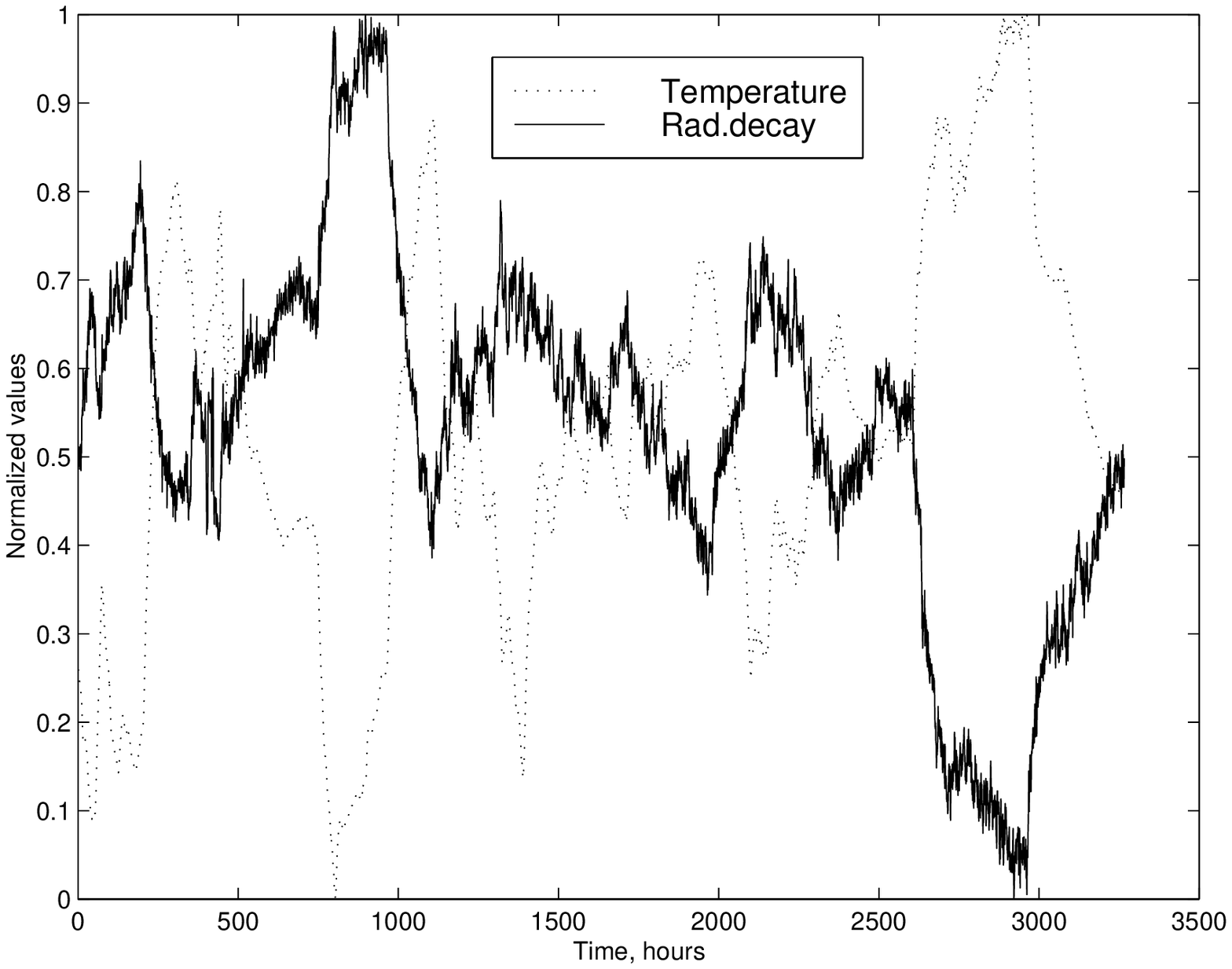,angle=0,height=60mm,width=60mm}}
\vspace{5mm}
\centerline{Fig.6.}
\end{figure}

\begin{figure}[hb]  
\centerline{\psfig{figure=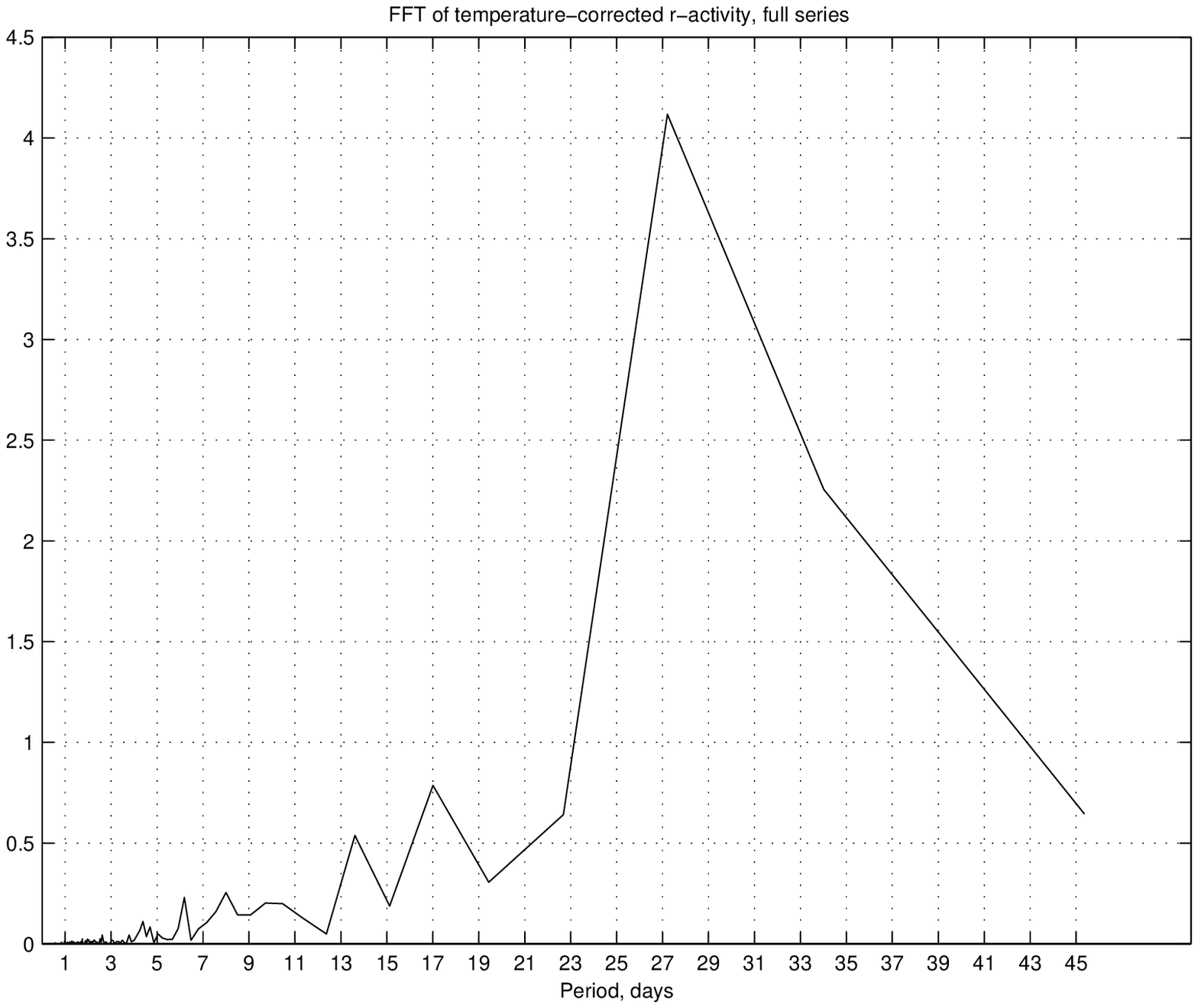,angle=0,height=60mm,width=60mm}}
\vspace{5mm}
\centerline{Fig.7.}
\end{figure}

\begin{figure}[hb]  
\centerline{\psfig{figure=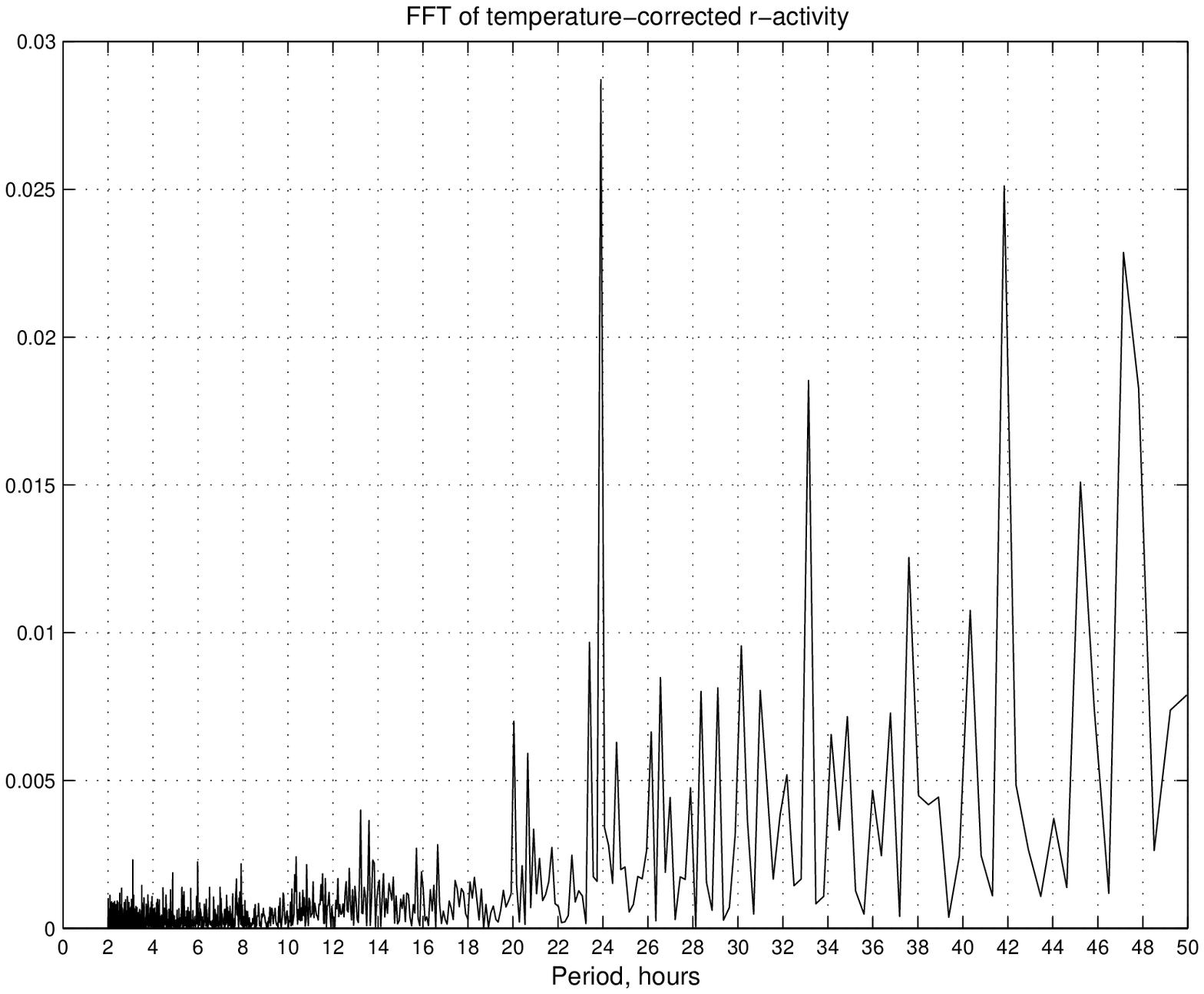,angle=0,height=60mm,width=60mm}}
\vspace{5mm}
\centerline{Fig.8.}
\end{figure}

\begin{figure}[hb]  
\centerline{\psfig{figure=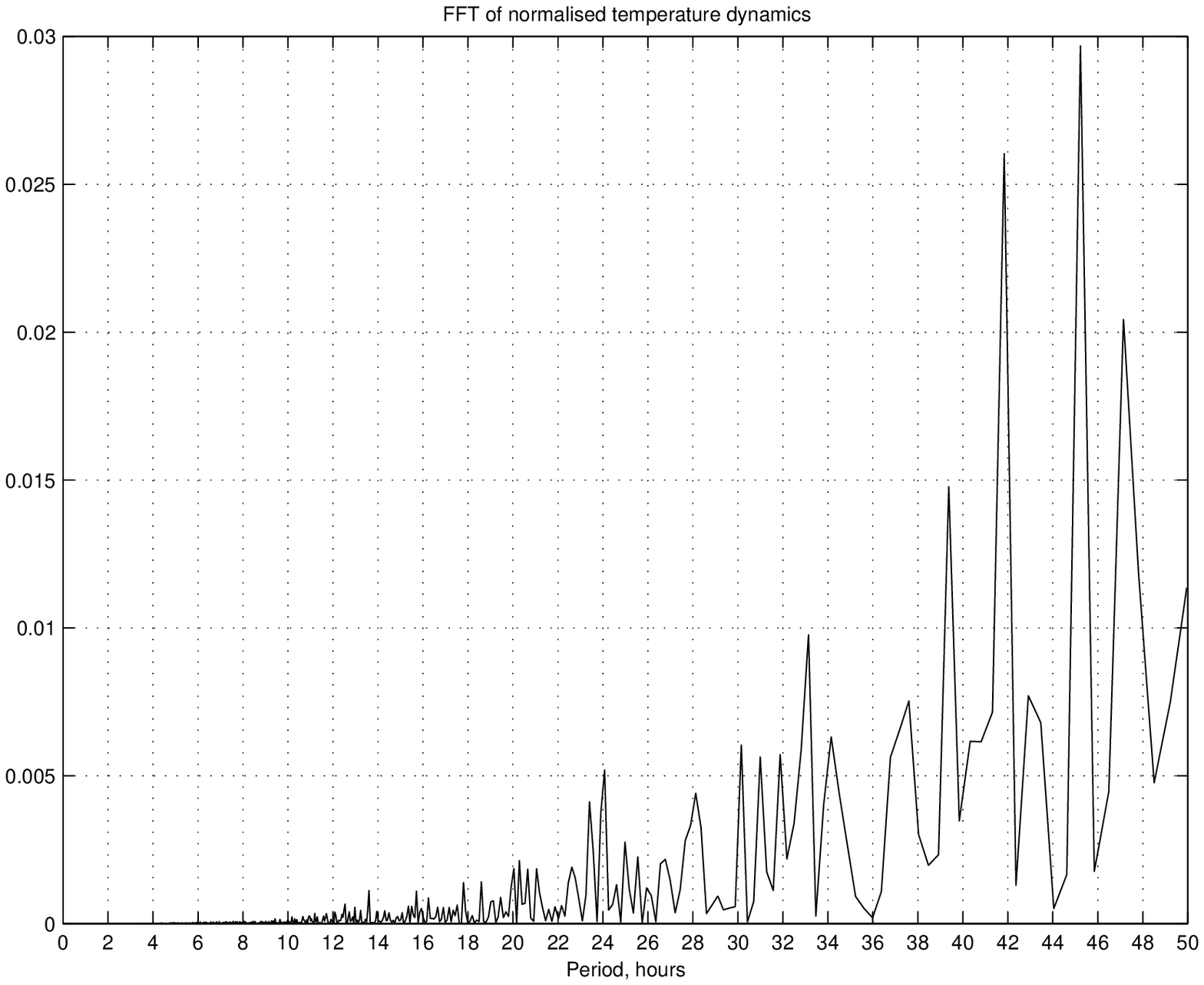,angle=0,height=60mm,width=60mm}}
\vspace{5mm}
\centerline{Fig.9.}
\end{figure}

\begin{figure}[hb]  
\centerline{\psfig{figure=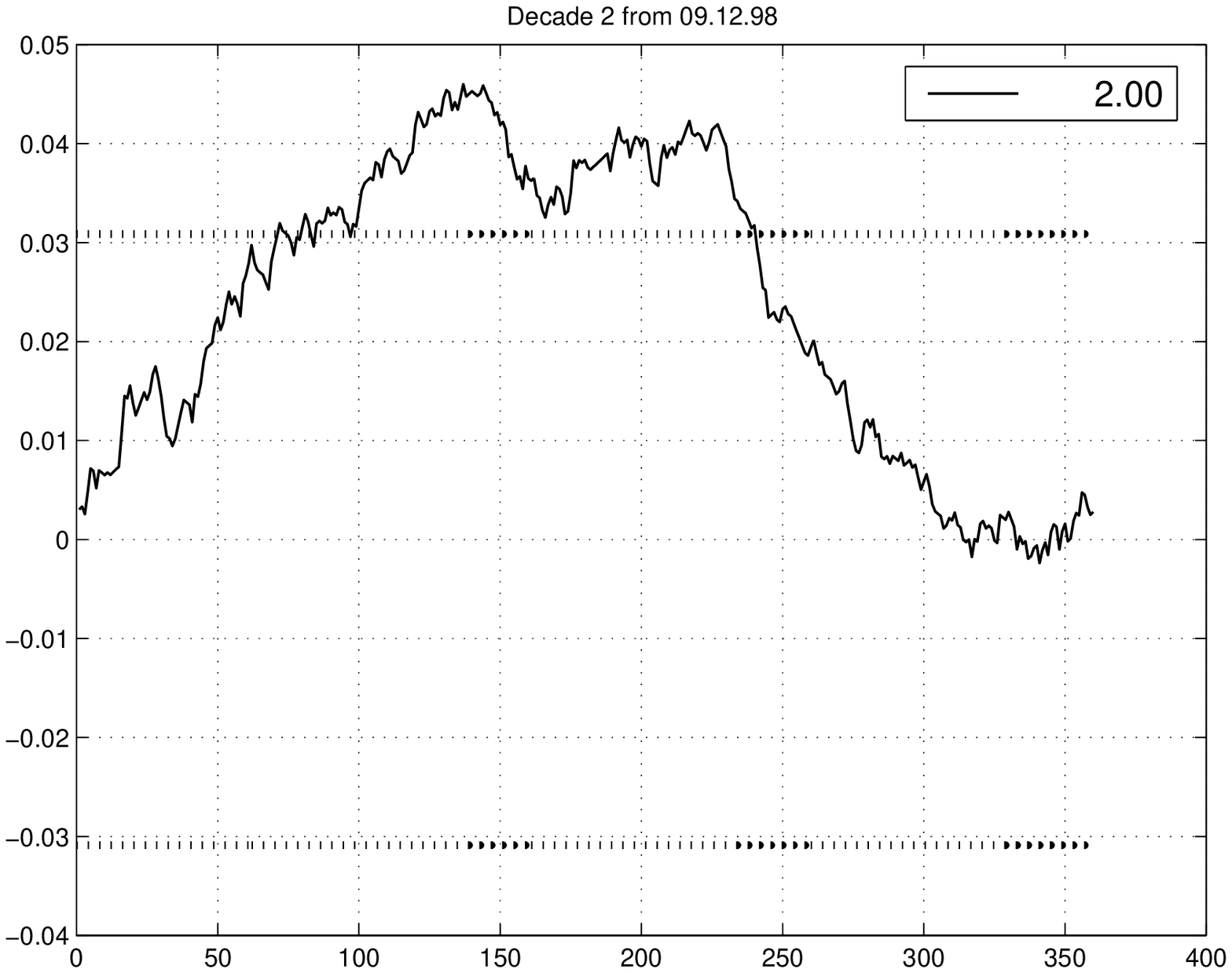,angle=0,height=60mm,width=60mm}}
\vspace{5mm}
\centerline{Fig.10.}
\end{figure}

\begin{figure}[hb]  
\centerline{\psfig{figure=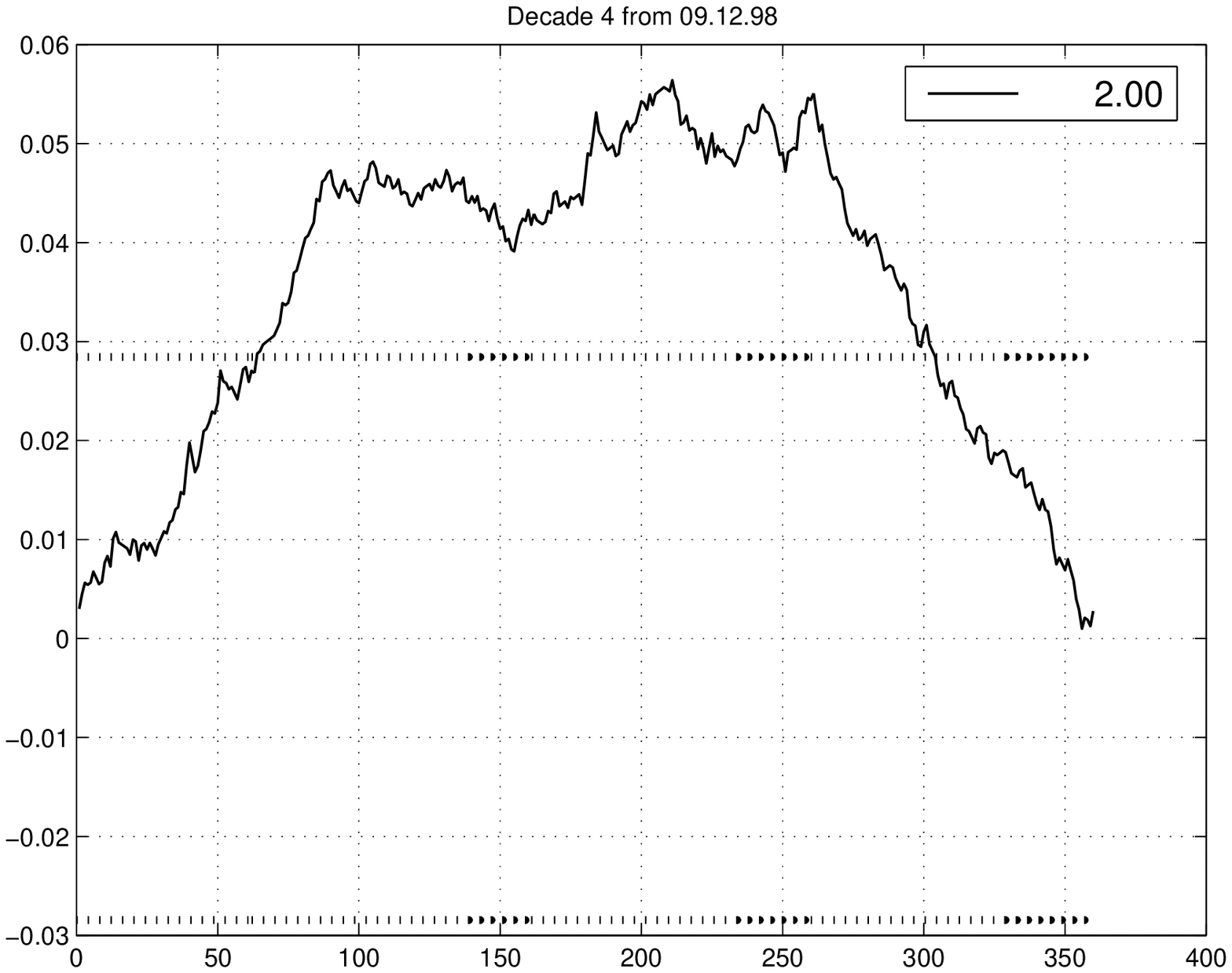,angle=0,height=60mm,width=60mm}}
\vspace{5mm}
\centerline{Fig.11.}
\end{figure}
\end{document}